# Improved Packet Forwarding Approach in Vehicular Ad hoc Networks using RDGR Algorithm

K.Prasanth[1] Dr.K.Duraiswamy[2] K.Jayasudha[3] and Dr.C.Chandrasekar[4]

[1]Research Scholar, Department of Information Technology,
K.S.Rangasamy College of Technology, Tamilnadu, India
prasanthkaliannan@gmail.com
[2]Dean Academic, Department of Computer Science,
K.S.Rangasamy College of Technology, Tamilnadu, India
drkduraiswamy@yahoo.com
[3]Research Scholar, Department of MCA,
K.S.R College of Engineering, Tamilnadu, India
jayasudhakaliannan@yahoo.com
[4]Reader, Periyar University, Tamilnadu, India
ccsekar@gmail.com

## ABSTRACT

*VANETs (Vehicular Ad hoc Networks) are highly mobile wireless ad hoc networks and will play an important role in public safety communications and commercial applications. Routing of data in VANETs is a challenging task due to rapidly changing topology and high speed mobility of vehicles. Position based routing protocols are becoming popular due to advancement and availability of GPS devices. One of the critical issues of VANETs are frequent path disruptions caused by high speed mobility of vehicle that leads to broken links which results in low throughput and high overhead . This paper argues the use of information on vehicles' movement information (e.g., position, direction, speed of vehicles) to predict a possible link-breakage event prior to its occurrence. So in this paper we propose a Reliable Directional Greedy routing (RDGR), a reliable position based routing approach which obtains position, speed and direction of its neighboring nodes from GPS. This approach incorporates potential score based strategy, which calculates link stability between neighbor nodes in distributed fashion for reliable forwarding of data packet.*

## 1. INTRODUCTION

Recent advances in wireless technologies have made inter-vehicular communications (IVC) possible in mobile ad hoc networks (MANETs) and this has given birth to a new type of MANET known as the vehicular ad hoc networks (VANETs). Internetworking over VANETs has been gaining a great deal of momentum over the past few years. VANETs is a form of mobile ad hoc network providing communications among nearby vehicles as well as between vehicles and nearby fixed equipment, usually described as roadside equipment. Vehicles are becoming "computer networks on wheels" and acts as mobile nodes of the network. VANET technology integrates ad hoc network, wireless LAN (WLAN) and cellular technology to achieve intelligent Inter-Vehicle Communications (IVC) and Roadside-to-Vehicle Communications (RVC). VANETs are a special case of MANETs and both are characterized by the movement and self-organization of the nodes. However, unlike MANETs, the mobility of vehicles in VANETs is, in general, constrained by predefined roads. Vehicle velocities are also restricted according to speed limits, level of congestion in roads, and traffic control mechanisms. In addition, given the fact that future vehicles can be equipped with devices with potentially longer transmission ranges, rechargeable source of energy, and extensive onboard storage capacities, processing power and storage efficiency are not an issue in VANETs as they are in MANETs. From these features, VANETs are considered as an extremely flexible and relatively "easy-to-manage" network pattern of MANETs.

Due to recent developments in the VANET field, a number of attractive applications, which are unique for the vehicular setting, have emerged. VANET applications include onboard active safety systems that are used to assist drivers in avoiding collisions and to coordinate among them at critical points such as intersections and highway entries. It is beneficial in providing intelligent transportation system (ITS) as well as drivers and passenger's assistant services. Safety systems may intelligently disseminate road information, such as incidents, real-time traffic congestion, high-speed tolling, or surface condition to vehicles in the vicinity of the subjected sites. This helps to avoid platoon vehicles and to accordingly improve road capacity. With such active safety systems, the number of car accidents and associated damage are expected to be largely reduced. In addition to the aforementioned safety applications, IVC communications can also be used to provide comfort applications. The latter may include weather information, gas station or restaurant locations, mobile e-commerce, infotainment applications, and interactive communications such as Internet access, music downloads, and content delivery.

VANETs have similar or different radio interface technologies, employing short-range to medium-range communication systems. The radio range of VANETs is several hundred meters, typically between 250 and 300 meters. In US, the Federal Communications Commission (FCC) has allocated 75 MHz in 5.9 GHz band for licensed Dedicated Short Range Communication (DSRC) for vehicle-to-vehicle and vehicle to infrastructure communications. Recently, the promises of wireless communications to support vehicular applications have led to several research projects around world. National Highway Traffic Safety Administration (NHTSA) and the automotive OEMs created the Vehicle Safety Communication Consortium (VSCC) to promote V2V networking for safety. Governments and prominent industrial corporations, such as Toyota, BMW, and Daimler–Chrysler, have launched important projects for IVC communications. Advanced Driver Assistance Systems (ADASE2) [1], Crash Avoidance Metrics Partnership (CAMP) [2], Chauffeur in EU [3],CarTALK2000 [4], FleetNet [5], California Partners for Advanced Transit and Highways (California PATH) [6],DEMO 2000 by Japan Automobile Research Institute (JSK), Electronic Toll Collection service (ETC), Advanced Cruise-Assist Highway System (AHS), Vehicle Information and Communication System (VICS) [7], AutoNet [8], Path [9], C2C-CC project [10] in Europe, and the related projects include eSafety Support [11], PReVENT project [12], Network on Wheels project [13], COMeSafety [14] etc are few notable examples. The Internet ITS (Intelligent Transportation Systems) Consortium [15] in Japan is one of the samples of VANETs projects. These projects are a major step toward the realization of intelligent transport services.

The design of effective vehicular communications poses a series of technical challenges. Guaranteeing a stable and reliable routing mechanism over VANETs is an important step toward the realization of effective vehicular communications. Existing routing protocols, which are traditionally designed for MANET, do not make use of the unique characteristics of VANETs and are not suitable for vehicle-to-vehicle communications over VANETs. Indeed, the control messages in reactive protocols and route update timers in proactive protocols are not used to anticipate link breakage. They solely indicate presence or absence of a route to a given node. Consequently, the route maintenance process in both protocol types is initiated only after a link-breakage event takes place. When a path breaks, not only portions of data packets are lost, but also in many cases, there is a significant delay in establishing a new path. This delay depends on whether another valid path already exists (in the case of multipath routing protocols) or whether a new route-discovery process needs to take place. The latter scenario introduces yet another problem. In addition to the delay in discovering new paths, flooding required for path discovery would greatly degrade the throughput of the network as it introduces a large amount of network traffic. In a highly mobile system such as VANET, where link breakage is frequent, flooding requests would largely degrade the system performance due to the introduction of additional network traffic into the system and interruption in data transmission.

## 2. RELATED RESEARCH

In this section, we briefly summarize the characteristics of VANETs related to routing and also we will survey the existing routing schemes in both MANETs and VANETs in vehicular environments.

### 2.1. VANETs Characteristics

In the following, we only summarize the uniqueness related to routing of VANETs compared with MANETs.

**Unlimited transmission power:** Mobile device power issues are not a significant constraint in vehicular Networks. Since the vehicle itself can provide continuous power to computing and communication devices.

**High computational capability:** Operating vehicles can afford significant computing, communication and sensing capabilities.

**Highly dynamic topology:** Vehicular network scenarios are very different from classic ad hoc networks. In VANETs, vehicles can move fast. It can join and leave the network much more frequently than MANETs. Since the radio range is small compared with the high speed of vehicles (typically, the radio range is only 250 meters while the speed for vehicles in freeway will be 30m/s). This indicates the topology in VANETs changes much more frequently.

**Predicable Mobility:** Unlike classic mobile ad hoc networks, where it is hard to predict the nodes' mobility, vehicles tend to have very predictable movements that are (usually) limited to roadways. The movement of nodes in VANETs is constrained by the layout of roads. Roadway information is often available from positioning systems and map based technologies such as GPS. Each pair of nodes can communicate directly when they are within the radio range.

**Potentially large scale:** Unlike most ad hoc networks studied in the literature that usually assume a limited network size, vehicular networks can in principle extend over the entire road network and so include many participants.

**Partitioned network:** Vehicular networks will be frequently partitioned. The dynamic nature of traffic may result in large inter vehicle gaps in sparsely populated scenarios and hence in several isolated clusters of nodes.

**Network connectivity:** The degree to which the network is connected is highly dependent on two factors: the range of wireless links and the fraction of participant vehicles, where only a fraction of vehicles on the road could be equipped with wireless interfaces.

### 2.2. Routing protocols in MANET

The routing protocols in MANETs can be classified by their properties. They can all be classified as either proactive, reactive, or hybrid.

**Proactive routing protocols** maintain and update information on routing between all nodes of a given network at all times. Route updates are periodically performed regardless of network load, bandwidth constraints, and network size. Routing information is stored in a variety of tables and are based on received control traffic. Generation of control messages and route calculation are driven by the routing tables. The main characteristic of proactive protocols is that nodes maintain a constantly updated understanding of the network topology. Consequently, a route to any node in the network is always available regardless of whether it is needed or not. While periodic updates of routing tables result in substantial signalling overhead, immediate retrieval of routes overcomes the issue of the initial route establishment delay in case of reactive protocols. Some of the protocols that have achieved prominence in the proactive category include optimized link state routing [16], hazy-sighted link state routing [17], topology broadcast based on reverse path forwarding [18], and destination-sequenced distance vector [19].

**Reactive routing protocols**, which are the flip-side of proactive protocols, route determination is invoked on a demand or need basis. Thus, if a node wishes to initiate communication with another host to which it has no route, a global search procedure is employed. This route-search operation is based on classical flooding search algorithms. Indeed, an RREQ message is generated and flooded, sometimes in a limited way, to other nodes. When the RREQ message reaches either the destination or an intermediate node with a valid route entry to the destination, a route-reply (RREP) message is sent back to the originator of the RREQ. A route is then set up between the source and the destination. Reactive protocols then remain passive until the established route becomes invalid or lost. Link breakage is reported to the source via a Route Error (RERR) message. Several protocols fall in this category. Notable examples are ad hoc on-demand distance vector (AODV) [20] and dynamic source routing (DSR) [21].

**Hybrid routing protocols** combine both the proactive and reactive approaches. Zone routing protocol (ZRP) is a notable example [22]. ZRP divides the network topology into different zones. Routing within zones, "intrazone routing," is performed by a proactive protocol. This yields no initial delay for routing among nodes from the same zone. On the other hand, to increase system scalability, routing between zones, "interzone routing," is done by a reactive protocol. While the hybrid approaches present an efficient and scalable routing strategy for large-scale environments, a number of key issues remain unsolved, and their implementation has not accordingly gained that much popularity within the researchers' community. Compared to reactive approaches, proactive protocols are easier to implement and exhibit relative stability. However, by applying them to a highly mobile environment such as VANETs, a storm of control messages is required to maintain an accurate view of the network topology. This intuitively results in heavy traffic contention, collisions of packets due to mass flooding broadcasts between neighbouring nodes, and, consequently, a significant waste of the scarce wireless bandwidth. They can be used only for environments where mobility is relatively static. Reactive protocols are thus preferred for dynamically changing environments where nodes have a few number of active routes (e.g., VANET) [23]. For a qualitative comparison between reactive and proactive schemes, the interested reader is referred to [24].

## 2.3. Routing protocols in VANET

Following are a summary of representative VANETs routing algorithms.

Traditionally, reactive protocols do not take into account mobility parameters during route discovery, resulting in paths which often break in highly mobile scenarios such as VANETs, causing excessive broadcasting and flooding the entire network in order for new routes to be discovered. Furthermore, the additional initial latency introduced by the route-discovery procedure poses serious challenges for reactive protocols. For this reason, reactive protocols, in their current format, are seen as inappropriate for time-critical applications such as cooperative collision avoidance (CCA), which is an important application type for vehicular communications. To cope with flooding, LAR [25], like other broadcast/flood reducing mechanisms [26], [27], directs broadcasting toward the estimated destination node. In [28], broadcast flood is limited only by forwarding consecutive RREQ packets which have a path hop accumulation smaller than the previous identical or duplicate RREQ packet. Otherwise, the newly arrived RREQ packet is dropped and hence not forwarded. Although these methods are quite satisfactory in providing efficient rebroadcasting with regard to coverage, integrating this broadcast minimizing schemes in routing does not consider path stability during the rebroadcasting procedure. Hence, we need a scheme that takes these issues into consideration, while reducing broadcast overhead. Attempts at predicting and selecting stable links have been proposed in [29]–[31]. However, they all depend on statistical analysis and probabilistic models of link duration.

The routing algorithm that considers stability in the routing criterion is the associativity based routing [32]. ABR uses associativity "ticks" messages (TICKs), which are periodically broadcasted in order to estimate the lifetime of links. If a node has high associativity ticks with its neighbour node, then the degree of stability (and hence link duration) is high. The destination node chooses nodes which have a high degree of associativity. If we consider ABR in a highly mobile pseudo linear mobile environment with no pause time, such as a VANET network or an aeronautical ad hoc network as introduced in [33], all nodes within a time range would receive equal associativity ticks regardless of their speed and direction. In this case, high associativity means that the neighbour node has been within range for a considerable period of time. It does not ensure that the mobile node will continue to remain within range, as the mobile node may already be close to the edge of the communication boundary. A better node which provides a more stable link may have just come into the range of the target node and would consequently have a lower associativity value. Thus, ABR would not be suitable for the considered mobility model.

Based from the aforementioned routing concepts, a set of routing protocols has been proposed for vehicular communications. While it is all but impossible to come up with a routing approach that can be suitable for all VANET applications and can efficiently handle all their inherent characteristics, attempts have been made to develop some routing protocols specifically designed for particular applications. For safety applications, a broadcast-oriented packet forwarding mechanism with implicit acknowledgment is proposed for intraplatoon CCA [34].In [35], a swarming protocol based on gossip messages is proposed for content delivery in future vehicular networks. For the provision of comfort applications, a segment-oriented data abstraction and dissemination (SODAD) is proposed in [36]. SODAD is used to create a scalable decentralized information system by local distribution of the information in vehicular networks. CarNet proposes a scalable routing system that uses geographic forwarding and a scalable distributed location service to route packets from vehicle to vehicle without flooding the network [37].

To avoid link rupture during data transmission, a movement-prediction-based routing (MOPR) is proposed in [38]. MOPR predicts future positions of vehicles and estimates the time needed for the transmission of data to decide whether a route is likely to be broken or not during the transmission time. The performance of the scheme largely depends on the prediction accuracy and the estimate of the transmission time that depends, in turn, on several factors such as network congestion status, driver's behaviour, and the used transmission protocols. In [39], a distributed movement-based routing algorithm is proposed for VANETs. This algorithm exploits the position and direction of movement of vehicles. The metric used in this protocol is a linear combination of the number of hops and a target functional, which can independently be calculated by each node. This function depends on the distance of the forwarding car from the line connecting the source and destination and on the vehicle's movement direction. Each vehicle needs to implement this in a distributed manner. DGRP is a position based greedy routing protocol [40], which uses the location, speed and direction of motion of their neighbors to select the most appropriate next forwarding node. Like GPSR it uses the two forwarding strategies greedy and perimeter. It predicts the position of nodes within the beacon interval whenever it needs to forward a data packet. This prediction can be done using previous known position, speed, and direction of motion of node. The weak link stability between the forwarding node and its neighbour node creates possibility of packet loss in DGRP. In highly mobile network, inaccurate position information leads to low throughput and high overhead. The frequent path disruptions caused by high speed mobility of vehicle that leads to broken links which results in low throughput and high overhead in DGRP.

## 3. PROPOSED WORK

### 3.1 Reliable Directional Greedy Routing (RDGR):

RDGR is a reliable position based greedy routing approach which uses the position, speed, direction of motion and link stability of their neighbours to select the most appropriate next forwarding node. It obtains position, speed and direction of its neighbouring nodes from GPS. If neighbour with most forward progress towards destination node has high speed, in comparison with source node or intermediate packet forwarder node, then packet loss probability is increased. In order to improve DGR protocol and increase its reliability, the proposed strategy introduces some new metrics to avoid loss of packets. The packet sender or forwarder node, selects neighbour nodes which have forward progress towards destination node using velocity vector, and checks link stability of those nodes. Finally, it selects one of them which has more link stability and sends packet to it. It uses combination metrics of distance, velocity, direction and link stability to decide about to which neighbour the given packet should be forwarded. Unlike DGR this approach not only uses the one hop neighbour's position, speed and direction of motion information, it also considers all neighbours position, speed, direction of motion information and link stability. This routing approach incorporates potential score based strategy, which reduces link breaks, enhances reliability of the route and improves packet delivery ratio.

### 3.2 Assumptions

The algorithm design is based on the following assumptions: All nodes are equipped with GPS receivers, digital maps, optional sensors and On Board Units (OBU). Location information of all vehicles/nodes can be identified with the help of GPS receivers. The only communications paths available are via the ad-hoc network and there is no other communication infrastructure. Node power is not the limiting factor for the design. Communications are message oriented. The Maximum Transmission Range (MTR) of each node in the environment is 250m.

### 3.3 Reckoning Link Stability:

To identify path stability we need to know individual link stability along the path. We define link stability in terms of link expiration time which means maximum time of connectivity between any two neighbor nodes. In order to calculate the link expiration time we assume motion parameters of any two neighbors are known. Let $n_1$ and $n_2$ be two nodes within the transmission range $R$ and $x_1'$, $y_1'$ and $x_2'$, $y_2'$ be the coordinate for node $n_1$ and $n_2$ with velocity $v_1$ and $v_2$ and direction $\theta_1$ and $\theta_2$ respectively. Let after a time interval $t$ the new coordinate will be $x_1$, $y_1$ for $n_1$ and $x_2$, $y_2$ for $n_2$. For time $t$ let $d_1$ and $d_2$ be the distance traveled by node $n_1$ and $n_2$.

$$d_1 = v_1 t$$
$$d_2 = v_2 t$$

Figure.1: Formula for Calculating d1 and d2

$$x_1 = x_1^i + x_1 = x_1^i + d_1 \cos \theta_1 = x_1^i + t(v_1 \cos \theta_1)$$

$$y_1 = y_1^i + y_1 = y_1^i + d_1 \sin \theta_1 = x_1^i + t(v_1 \sin \theta_1)$$

$$x_2 = x_2^i + x_2 = x_2^i + d_2 \cos \theta_2 = x_2^i + t(v_2 \cos \theta_2)$$

$$y_2 = y_2^i + y_2 = y_2^i + d_2 \sin \theta_2 = y_2^i + t(v_2 \sin \theta_2)$$

Figure.2: Formula for calculating new coordinates (with respect to old coordinates)

$$D^2 = \{(x_1^i - x_2^i) + t(v_1 \cos \theta_1 - v_2 \cos \theta_2)\}^2 + \{(y_1^i - y_2^i) + t(v_1 \sin \theta_1 - v_2 \sin \theta_2)\}^2$$

$$D = \sqrt{\begin{array}{c}\{(x_1^i - x_2^i) + t(v_1 \cos \theta_1 - v_2 \cos \theta_2)\}^2 \\ + \{(y_1^i - y_2^i) + t(v_1 \sin \theta_1 - v_2 \sin \theta_2)\}^2\end{array}}$$

Figure.3: Formula for calculating distance between two nodes at time t

$$LS = \frac{R}{D} = \frac{R}{\sqrt{\begin{array}{c}\{(x_1^i - x_2^i) + t(v_1 \cos \theta_1 - v_2 \cos \theta_2)\}^2 \\ \{(y_1^i - y_2^i) + t(v_1 \sin \theta_1 - v_2 \sin \theta_2)\}^2\end{array}}}$$

*Here*,
**LS** : *link stability between any two nodes over time period t.*
**R** : *Maximum transmission range.*
**D**:*Distance between two nodes at time t.*

Figure.4: Formula for calculating link stability between two nodes at time t

We can calculate $d_1$ and $d_2$ using the following formula shown in Figure.1. The new coordinates (with respect to old coordinates) can be calculated using the formula shown in Figure.2. The distance between two nodes at time t will be obtained from the formula shown in Figure.3.When the distance between two nodes becomes larger than the transmission range the nodes will be disconnected. For transmission range $R$ link stability $LS$ between any two nodes over time period $t$ can be calculated by the formula shown in Figure.4. Note that $LS$ is the link stability of individual links between any two nodes.

### 3.4 Potential Score Calculation:

The potential score (PS) of all nodes present within the transmission range of source/packet forwarding node is calculated. The potential score (PS) is calculated to identify the closeness of next hop to destination, direction of motion of nodes and reliability of neighbor nodes. The appropriate node with largest potential score will be considered as having higher potential to reach the destination node and that particular node can be chosen as next hop to forward the packet to the destination node. The potential score of neighboring nodes can be calculated by the following mathematical model represented in Figure.5. The neighbor node with high potential score is considered as the most suitable next hop for forwarding packet with stable link. The RDGR algorithm is explained in Figure.6.

$$Potential\ score(PS_i) = \alpha \times \left(1 - \frac{D_i}{D_c}\right) + \beta \times \cos(\overline{v}_i, \overline{l}_{i,d}) + \lambda \times LS_{c,i}$$

Here,

$PS_i$ : Potential score of node i

$\rho, \beta$ : Two potential factors

Let $\rho + \omega = 1$ and $\rho > \omega$

$D_i$ : Shortest distance from neighbor node i to destination D.

$D_c$ : Shortest distance from packet forwarding node c to destination D.

$\frac{D_i}{D_c}$ : Closeness of nexthop.

$\overline{v}_i$ : Vector for velocity of edge node i.

$\overline{l}_{i,d}$ : Vector for the location of edge node i to the location of destination node D

$\cos(\overline{v}_i, \overline{l}_{i,d})$ : Cosine value of angle made by these vectors

$LS_{c,i}$ : Link Stability between packet forwarding node c to neighbor node i

Figure.5: Potential Score Calculation

$MTR$: Maximum Transmission Range $= 250m$

$currentnode$: the current packet carrier

$loc_c$: the location of current node

$\overline{v_c}$: speed vector for current node

$dest$: destination of the packet

$loc_d$: the location for destination

$nextHop$: the node selected as next hop

$neigh_i$: the ith neighbor

$loc_i$: the location of the ith neighbor

$\overline{v}_i$: the speed vector of the ith neighbor

1.  $loc_c \leftarrow getLocation(currentnode)$
2.  $\overline{v_c} \leftarrow getSpeed(currentnode)$
3.  $loc_d \leftarrow getLocation(dest)$
4.  $D_c = distance(loc_c, loc_d)$
5.  $\overline{l}_{c,d} = loc_d - loc_c$
6.  $PS = \omega \times \cos(\overline{v_c}, \overline{l}_{c,d})$
7.  $nextHop = currentnode$
8.  **for all** neighbors of currentnode **do**
9.  $loc_i \leftarrow getLocation(neigh_i)$
10. $\overline{v}_i \leftarrow getSpeed(neigh_i)$
11. $D_i = distance(loc_d, loc_i)$
12. $D_{ci} = distance(loc_c, loc_i)$
13.      **for all** neighbors of currentnode with $D_{ci}$ **do**
14. **if** $(D_{ci} < MTR)$
15. $\overline{l}_{i,d} = loc_d - loc_i$
16. $PS_i = \rho \times \left(1 - \frac{D_i}{D_c}\right) + \omega \times \cos(\overline{v}_i, \overline{l}_{i,d}) + \lambda \times LS_{c,i}$
17. **for** $neigh_i$ with greater $PS_i$ **do**
18. $PS = PS_i$
19. $nextHop = neigh_i$
20. **end for**
21. **else**
22. carry the packet with currentnode
23. **end if**
24. **end for**
25. **end for**

Figure.6: Pseudo code of RDGR Algorithm

## 4. SIMULATION RESULTS

In this section, we evaluate the performance of routing protocols of vehicular networks in an open environment. So among the routing protocols we aforementioned, we choose GPSR, DGRP and RDGR for comparison.

### 4.1 Revival Mobility model (RMM)

We use Revival Mobility model (RMM) to simulate the movement pattern of moving vehicles on streets or roads defined by maps from the GPS equipped in the vehicles. In Revival Mobility model (RMM), the road comprises of two or more lanes. Vehicles or nodes are randomly distributed with linear node density. Each vehicle can move in different speed. This mobility model allows the movement of vehicles in two directions. i.e. north/south for the vertical roads and east/west for the horizontal roads. In cross roads, vehicles choose desired direction based on the shortest path. A security distance should be maintained between two subsequent vehicles in a lane. Overtaking mechanism is applicable and one vehicle can able to overtake the preceding vehicle. Packet transmission is possible and can be done by vehicles moving in both directions, which means front hopping and back hopping of data packet is possible as shown in the Figure.7.

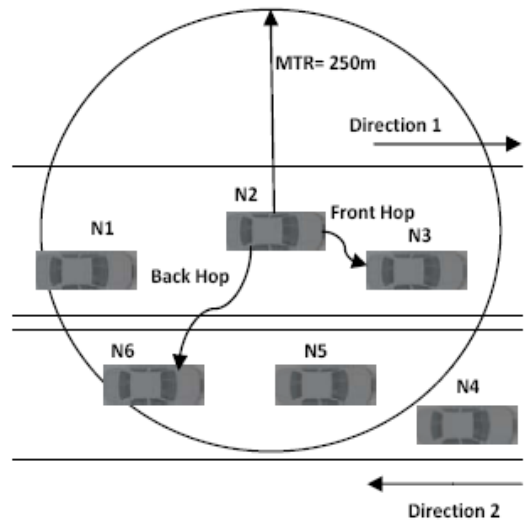

Figure.7 Revival Mobility Model

In this mobility, deterministic and instantaneous transmission mechanism in which a message is available for receiving within a certain radius $r=250m$ from the sender with certainty, but unavailable further away. Vehicles can unicast and broadcast packets to the neighbour vehicle which is present within its transmission range.

The Simulations were carried out using Network Simulator (NS-2) ([41]). We are simulating the vehicular ad hoc routing protocols using this simulator by varying the number of nodes. The IEEE 802.11 Distributed Coordination Function (DCF) is used as the Medium Access Control Protocol. The packet size was fixed to 512 bytes. The Traffic sources are UDP. Initially the nodes were placed at certain specific locations, and then the nodes move with varying speeds towards new locations. The nodes move with speeds up to 25 meter/sec. For fairness, identical mobility and traffic scenarios were used across the different simulations. The simulation parameters are specified in Table 1

**4.2 Performance Metrics:**

**Packet Delivery Ratio:** It is defined as the number of correctly received packets at the destination over the number of packets sent by the source.

$$PDR = \frac{Total\ number\ of\ packets\ delivered}{Total\ number\ of\ packets\ transferred} \times 100$$

Table1: Simulation Parameters

| Parameter | Value |
|---|---|
| Simulation Area | 1000m * 1000m |
| Number of Vehicles | 20 - 100 |
| Average speed of vehicles | 0 – 25 metre/second |
| Number of packet Senders | 40 |
| Transmission Range | 250m |
| Constant Bit Rate | 2 (Packets/Second) |
| Packet Size | 512 Bytes |
| Vehicle beacon interval | 0.5 (Seconds ) |
| MAC Protocol | 802.11 DCF |

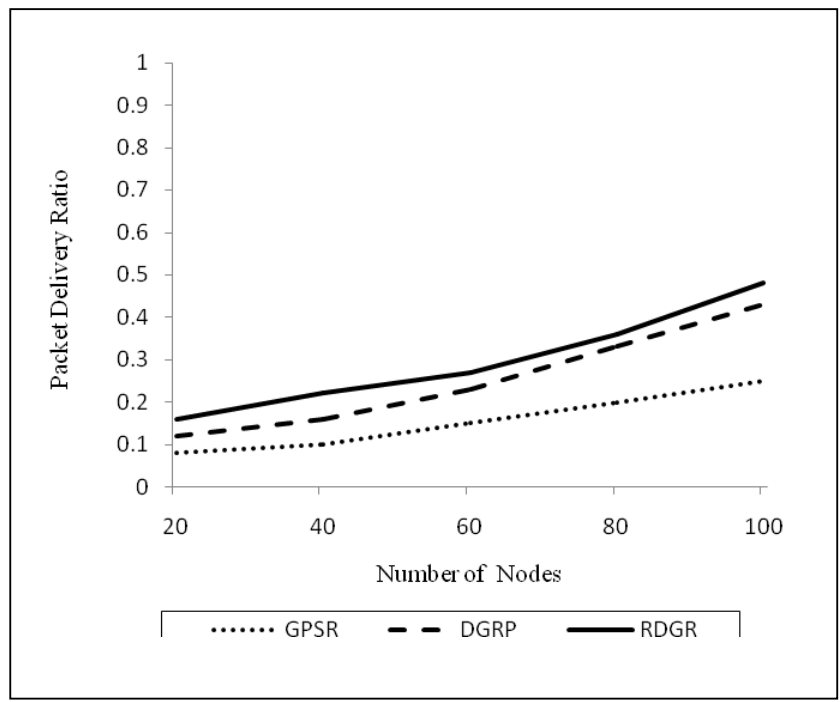

Figure.8 Packet delivery ratio Vs Number of nodes

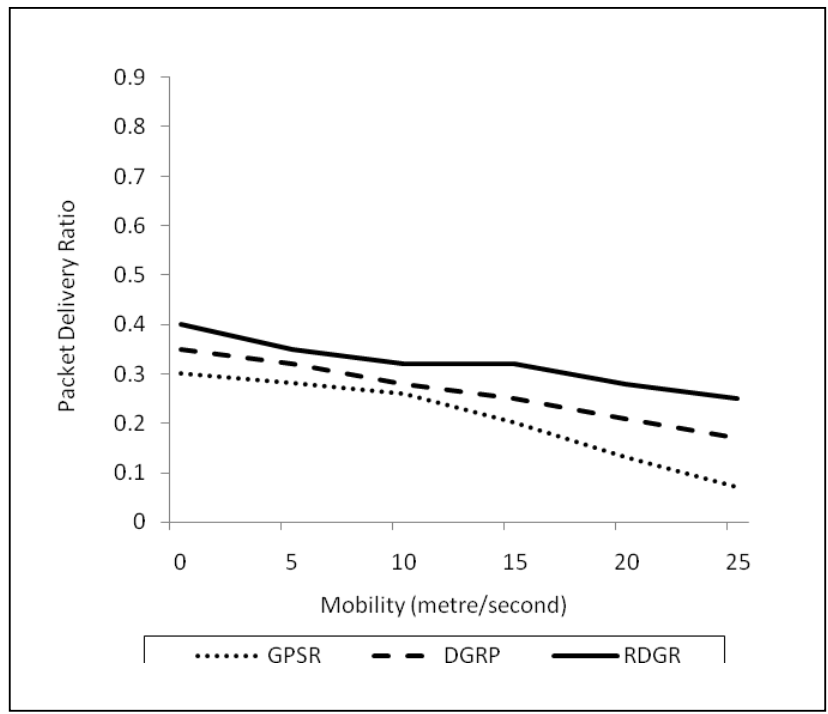

Figure.9 Packet delivery ratio Vs Mobility

We have simulated the proposed RDGR and compared it with GPSR and DGRP. In real network, each node finds its position by a positioning system like GPS, but in simulation we give every node, its position without any expense. In real network and in simulation every node propagates its position to its neighbors' periodically. In real network, the packet source node finds the location of destination node by a suitable location server (i.e. GPS). But in simulation, we give position of destination node, to source node, without any expense. To simulate the proposed strategy, sender node selects neighbor node which possess high potential score. We tested our approach with Revival Mobility Model with simulation parameters shown in the Table 1. We have considered the moving destination node in our simulation. We have evaluated GPSR, DGRP and RDGR for packet delivery ratio under different vehicle density and different mobility.

We can see from Figure.8, the performance metrics shows that gain in packet delivery ratio increases as number of node increases. This is because more number of nodes provides opportunity to select suitable neighbor node. In RDGR approach, a reliable neighbor node was chosen using potential score base strategy. At each instance of packet forwarding, the potential score is calculated for packet forwarding node and its neighbor nodes. This results in minimized packet loss and improved packet delivery during packet forwarding event. Our approach is comparatively better than GPSR and DGRP with average gain of packet delivery ratio and the overall packet delivery ratio is improved about 6% compared to DGRP.

We can see from Fig.9, the performance metrics shows that gain in packet delivery ratio decreases as speed increases. This is because with the increase in the speed, there is a possibility of decrease in link stability. In RDGR approach, a reliable neighbor node was chosen using potential score base strategy. At each instance of packet forwarding, the potential score is calculated for packet forwarding node and its neighbor nodes. So it reduces link breaks when the mobility is more than 20 m/s. The decrease in link breaks provides improve packet delivery ratio. Our approach is comparatively better than GPSR and DGRP with average gain of packet delivery ratio and the overall packet delivery ratio is improved about 8% over the speed of 5m/s to 25m/s.

## 5. CONCLUSION

In this paper we have investigated routing aspects of VANETs. We have identified the properties of VANETs and previous studies on routing in MANETs and VANETs. We have commented on their contributions, and limitations. By using the uniqueness of VANETs, we have proposed Revival Mobility Model and a new position based greedy routing approach RDGR. We have proposed a Reliable Directional greedy routing approach and it uses potential score based strategy, which calculates link stability between neighbour nodes for reliable forwarding of data packet. RDGR acquire position, direction and speed of neighbour nodes using GPS. The effect of link duration calculation was studied. We have simulated RDGR in ns-2 using Revival Mobility model. The results shown that our approach perform better in packet delivery ratio as compared to GPSR and DGRP. Another interesting problem is to study the behaviour of RDGR with various mobility models.

**Authors**

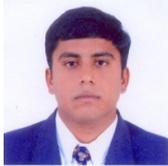

**K.Prasanth** received the B.E degree in Computer Science from K.S.Rangasamy College of Technology, Tiruchengode in 2005, .Tech in Computer Science from SRM University, Chennai in 2007, MBA System) in Periyar University, Salem in 2007. He worked as Project Engineer in Wipro Technologies, Bangalore from (2007-2008).He is currently working as lecturer in Department of Information Technology, K.S.Rangasamy College of Technology. His Current research interest includes Mobile Computing, Mobile Ad Hoc Networks and Vehicular Ad Hoc Networks

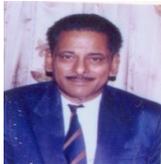

**Dr.K.Duraiswamy** received the B.E., M.Sc. and Ph.D. degrees, from the University of Madras and Anna University in 1965, 1968 and 1987 respectively. He worked as a Lecturer in the Department of Electrical Engineering in Government College of Engineering, Salem from 1968, as an Assistant professor in Government College of Technology, Coimbatore from 1983 and as the Principal at K.S.Rangasamy College of Technology from 1995. He is currently working as a Dean in the Department of Computer Science and Engineering at K.S.Rangasamy College of Technology (Autonomous Institution).His research interest includes Mobile Computing, Soft Computing, Computer Architecture and Data Mining. He is a senior member of ISTE, IEEE, and CSI.

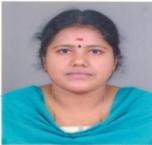

**K. Jayasudha** received the B.Sc degree in 2002, MCA degree in 2005, M.E degree in Computer Science in 2007, from K.S.Rangasamy College Of Technology, Thiruchengodu. She is currently working as Lecturer in Department Of Computer Applications, K.S.R.College Of Engineering, Thiruchengodu. Her current research interest includes Data Mining, Vehicular Networks.

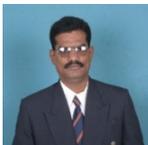

**Dr.C.Chandrasekar** received the B.Sc degree and M.C.A degree. He completed PhD in periyar university Salem at 2006. He worked as Head of the Department, Department Of Computer Applications at K.S.R.College of Engineering from 2007. He is currently working as Reader in the Department of Computer Science at Periyar University, Salem. His research interest includes Mobile computing, Networks, Image processing, Data mining. He is a senior member of ISTE, CSI.